\documentclass[12pt]{article}
\usepackage{epsfig,amsfonts,amssymb,amsbsy,amsbsy}
\def\cvp{\raise 2pt\hbox{,}}

\def\plb#1#2#3{{\it Phys.\ Lett.\ }{\bf B #1} (#2) #3}
\def\npb#1#2#3{{\it Nucl.\ Phys.\ }{\bf B #1} (#2) #3}
\def\prl#1#2#3{{\it Phys.\ Rev.\ Lett.\ }{\bf #1} (#2) #3}
\def\jhep#1#2#3{{\it J. High Energy Phys.\ }{\bf #1} (#2) #3}
\def\prd#1#2#3{{\it Phys.\ Rev.\ }{\bf D #1} (#2) #3}
\def\atmp#1#2#3{{\it Adv.\ Theor.\ Math.\ Phys.\ }{\bf #1} (#2) #3}
\def\cmp#1#2#3{{\it Comm.\ Math.\ Phys.\ }{\bf #1} (#2) #3}
\def\pr#1#2#3{{\it Phys.\ Rep.\ }{\bf #1} (#2) #3}

\def\ijmpa#1#2#3{{\it Int.\ J.\ Mod.\ Phys.\ }{\bf A #1} (#2) #3}
\def\aam#1#2#3{{\it Adv. in Appl. Math.\ }{\bf #1} (#2) #3}
\def\suN{{\rm SU}(N)}

\def\rc{r_{\rm c}}\def\gc{g_{\rm c}}
\begin{document}
%
%
\pagestyle{empty}
{\parskip 0in
\hfill PUPT-1993

\hfill LPTENS-01/10

\hfill hep-th/0107096}

\vfill
\begin{center}
{\large\bf Exact amplitudes} \\
\medskip
{\large\bf in four dimensional non-critical string theories}

\vspace{0.4in}

Frank F{\scshape errari}\footnote{On leave of absence from Centre 
National de la Recherche Scientifique, Laboratoire de Physique 
Th\'eorique de l'\'Ecole Normale Sup\'erieure, Paris, France.}\\
\medskip
{\it Joseph Henry Laboratories\\
Princeton University, Princeton, New Jersey 08544, USA}\\
\smallskip
{\tt fferrari@feynman.princeton.edu}
\end{center}

\vfill\noindent
The large $N$ expansion of ${\cal N}=2$ supersymmetric Yang-Mills theory 
with gauge group $\suN$ has 
recently been shown to break down at singularities on the moduli space. 
We conjecture that by taking $N\rightarrow\infty$ and approaching the
singularities in a correlated way, all the observables of the theory have
a finite universal limit yielding amplitudes in string theories dual to
field theories describing the light degrees of freedom.
We explicitly calculate the amplitudes corresponding to the
Seiberg-Witten period integrals for an $A_{n-1}$ series of 
multicritical points as well as for other critical points 
exhibiting a scaling reminiscent of the $c=1$ matrix model.
Our results extend the matrix model approach to non-critical 
strings in less than one dimension to non-critical strings in 
four dimensions.
\vfill
\begin{flushleft}
July 2001
\end{flushleft}
\newpage\pagestyle{plain}
\baselineskip 16pt
%
\section{Introduction}
Deriving exact results in string theory, or more modestly results
valid to all orders of perturbation theory, is a challenge that can be 
rarely taken up. The first real success in this direction was achieved in 
1989 \cite{BK} by using the relationship between the large $N$ expansion 
of matrix integrals and a discretized version of the sum over world sheets
\cite{tHooft,DK}. A crucial property is that the large $N$
expansion of a matrix theory can break down at some critical value 
$g_{\rm c}$ of the 't Hooft coupling $g$. For example, in simple
integrals over hermitian matrices \cite{IZ},
the partition function has an asymptotic expansion of the form
\begin{equation}
\label{Zexp}
Z(g) = \sum_{h\geq 0} N^{2 -2h}\, Z_h(g),
\end{equation}
where the coefficients $Z_h(g)$ are given themselves by a convergent
series in
$g$. The radius of convergence of this series, which turns out to be
independent of $h$, gives the critical
coupling $g_{\rm c}$. For $g$ near $g_{\rm c}$, the terms with a high
power of $g$, or equivalently the Feynman diagrams with a very large
number of vertices, dominate. In the dual double-line representation of
the graphs, such Feynman diagrams correspond to discretized oriented
Riemann surfaces with a very large number of faces, and thus
give a good approximation to the smooth world sheets of an oriented closed 
string theory. It is then possible to take a double scaling limit
$N\rightarrow \infty$ and $g\rightarrow g_{\rm c}$ from which a genuine
continuum string theory emerges \cite{BK}. 
The double scaled observables, which
are identified with amplitudes in non-critical string theories, 
can be explicitly studied in various cases corresponding to a world
sheet theory of central charge $c\leq 1$ (critical strings propagating in
less than two dimensions). For reviews and references on these
developments, the reader may consult \cite{review}.

In spite of its impressive achievements, the above approach suffers from
two main drawbacks that have never been satisfactorily solved. The first
disappointment is that the double
scaling limits do not provide a non-perturbative definition of unitary
theories. Technically, one can trace the problem to the fact that in 
the unitary cases the
critical coupling $g_{\rm c}$ either is negative (and thus the original
matrix integral is not convergent at $g=g_{\rm c}$), or does not
correspond to a true saddle point. All the information is encoded in
non-linear differential equations (like the Painlev\'e equation)
which yield recursion relations determining 
the perturbative expansions to all orders, but can
accomodate various non-perturbative extensions. In the case of
non-unitary models, the perturbative series are Borel summable and thus
a natural ansatz for a non-perturbative definition can be given, but
this is not possible for the most interesting unitary cases.
A second major difficulty is that the method can be used only on
tractable matrix integrals, which narrows drastically
the possible applications. Critical strings propagating 
in $D+1$ dimensions, or equivalently non-critical string 
theories in $D$ dimensions,
are a priori related to $D$ dimensional matrix path integrals near
critical points where the large $N$ expansion breaks down, of which no
example were known for $D>1$ in the 90s. Only strings propagating in 
less than two space-time dimensions (the so-called $c=1$ barrier) were
thus studied at that time. 

This situation has changed very recently with a work of the author on the
large $N$ limit of four dimensional
${\cal N}=2$ supersymmetric gauge theories \cite{fer}.
The matrix integrals one considers in this context are 
supersymmetrized versions of the gauge
theory path integrals that were originally discussed by 't~Hooft
\cite{tHooft}. Such path integrals need to be renormalized and the
't~Hooft coupling $g$ transmutes into a unit of mass and is not a free
parameter. The ``parameters''
that can be adjusted to critical values are in
these cases Higgs vaccuum expectation values (moduli).
One of the main result of \cite{fer} was to
show explicitly that the large $N$ limit does break
down at critical points (more commonly called singularities) on the
moduli space. The aim of the present work is to argue that double scaling
limits can be defined near those critical points, and to extract some
exact string
amplitudes from the double scaled Seiberg-Witten period integrals.
Since the gauge theory path integrals are non-perturbatively 
defined for all moduli, we always obtain a full 
non-perturbative prescription for the amplitudes, in sharp contrast to
the $c\leq 1$ cases.

Our main results were originally guessed by
studying simple two dimensional toy models \cite{fer2} whose main virtue
is that they are tractable both in supersymmetric and non-supersymmetric
examples. A supersymmetric version of the models, closely related to
${\cal N}=2$ super Yang-Mills \cite{Dorey}, turned out to be
particularly instructive
\cite{fer3}. The works \cite{fer2,fer3} themselves built on early attempts
to study double scaling limits for $D>1$ in
simple ${\rm O}(N)$ vector models \cite{ZJM}.

We have organized the paper as follows. In Section 2 we start by
giving an elementary
discussion of the simplest critical point. Then we state the general
conjecture and discuss its physical content in some details. We have tried
to give a comprehensive qualitative presentation of our main points, 
without entering into any detailed calculation. In Section 3 we 
define the scalings near various multicritical points and show
that the Seiberg-Witten period integrals have a finite limit. 
We give explicit formulas for the resulting integrals,
which correspond to
exact amplitudes in four dimensional non-critical string theories. Section 
4 is a short conclusion.
\vfill\eject

\section{A simple example and the general conjecture}
\subsection{The simplest critical point}

Let us start with the final formula of \cite{fer},
\begin{eqnarray}
\label{z1}
&&\hskip -1.75cm {i\pi z_1\over N} = 2\sqrt{1-1/r^{2}} - 2 \ln\left( 
1+\sqrt{1-1/r^{2}}\right) - 2\ln r - {2\sqrt{1-1/r^{2}}\,\ln 2\over N}
\nonumber\\
&&\hskip 6cm +{\pi^{2}/12 - (\ln 2)^{2}\over 
N^{2}r^{2}\sqrt{1-1/r^{2}}} +
{\cal O}\bigl(1/N^{3}\bigr).\\\nonumber
\end{eqnarray}
It gives the large $N$ expansion of a Seiberg-Witten period integral
$z_1$ in pure ${\cal N}=2$, $\suN$ super Yang-Mills theory, as a function
of a dimensionless modulus 
\begin{equation}
\label{rdef}
r = v/\Lambda\, ,
\end{equation}
where $v$ is the global mass scale
setting the Higgs eigenvalues distribution, and $\Lambda$
the dynamically generated scale of the theory. At $r=\rc=1$ there is a
singularity where a dyon is becoming massless. The mass of the dyon is
given by
\begin{equation}
\label{MBPS}
M_{\rm BPS} = \sqrt{2}\, | v\, z_1|.
\end{equation}
At leading $N=\infty$ order, the formula (\ref{z1}) incorrectly predicts 
$M_{\rm BPS}\propto (r-\rc)^{3/2}$ as $r\rightarrow\rc$.
The correct answer is known from electric/magnetic
duality \cite{SW} to be $M_{\rm BPS}\propto r-\rc$. Equation (\ref{z1}) 
also explicitly shows that higher order corrections are infinite at $r=\rc$. 
These infrared divergences were discussed at length in \cite{fer}.

Let us now introduce the deviation to the critical point
\begin{equation}
\label{delta}
\delta = r-\rc
\end{equation}
and consider the quantity $z_1/\sqrt{\delta}$ in the limit
$\delta\rightarrow 0$. By keeping only the dominant contribution
at each order in $1/N$, we get
\begin{equation}
\label{a1simple}
{i\pi z_1\over\sqrt{\delta}} = -{2\over 3}\, N\delta - 2\ln 2+ 
{\pi^2/12 - (\ln 2)^2\over N\delta} + {\cal O}(1/N^2).
\end{equation}
In the double scaling limit
\begin{equation}
\label{dsl1}
N\rightarrow\infty\, , \quad \delta\rightarrow 0\, ,\quad
N\delta = {\rm constant}\, ,
\end{equation}
we see that the first three terms in the expansion of
$z_1/\sqrt{\delta}$ remain finite.
Introducing the dimensionless coupling constant
\begin{equation}
\label{kappadef}
\kappa ^{-1} = N\delta
\end{equation}
and the rescaled amplitude
\begin{equation}
\label{a1def}
{\cal A}_1 = {i\pi z_1\over\sqrt{\delta}}
\end{equation}
we have 
\begin{equation}
\label{a1p}
{\cal A}_1 = -{2\over 3\kappa} - 2\ln 2 + 
\left(\pi^2/12 - (\ln 2)^2\right) \kappa + \cdots
\end{equation}
We want to interpret (\ref{a1p}) as giving the first terms in the
perturbative expansion of a string theory amplitude ${\cal A}_1$. The
full amplitude is given by the full Seiberg-Witten period $i\pi
z_1/\sqrt{\delta}$ in the scaling limit (\ref{dsl1}), which indeed yields
a finite result as we will show in Section 3. But before we embark upon 
those calculations, we are going to discuss the general ideas 
at the root of the identification of ${\cal A}_1$ and its
generalizations with string theory amplitudes.

\subsection{The conjecture}

The main claims of this paper can be summarized as follows.
Near singularities on the moduli space of ${\cal
N}=2$ supersymmetric Yang-Mills theory where the large $N$ expansion
breaks down \cite{fer}, it is possible to define double scaling limits,
similar to (\ref{dsl1}), with the following properties:\\
\noindent i) after a suitable rescaling of the space-time variables,
corresponding to a low energy limit, all the
correlation functions of the gauge theory have a finite limit.  
Most of the degrees of freedom of the original gauge 
theory decouple in the
scaling limit, and we are left with a double scaled field theory
describing the interacting light degrees of freedom only.\\
\noindent ii) the double scaling limits define $D=4$ continuum unitary
non-critical string theories dual to the double scaled field theory.

The ideas underlying
the conjecture are largely independent of supersymmetry, as the study of
two dimensional models clearly shows \cite{fer2}, and
similar statements could be made in the more general
context of four dimensional gauge theories with Higgs fields.
However, concrete examples are limited so far to ${\cal N}=2$ 
supersymmetric theories.

\subsection{Discussion}
\subsubsection{The double scaled gauge theory}
The ability to define the double scaled gauge theory near the critical 
points relies on the fact that the infrared divergences responsible for 
the breakdown of the $1/N$ expansion \cite{fer}
are specific enough so that they can
be compensated for by taking $N\rightarrow\infty$ in a suitable way. 
We will prove in Section 3 that this is possible for the low energy
observables for which Seiberg and Witten have provided
exact formulas \cite{SW}. The conjecture is then that all the other
correlators of the gauge theory will also have a finite limit. It is 
actually very important to realize that the double scaling limit is
always a low energy limit. This is not surprising, since the
terms that survive are the most divergent at
each order in $1/N$, and those divergences are infrared effects as
stressed in \cite{fer}. This is also the simplest way to understand 
the universality of the limit. For example, the amplitude ${\cal
A}_1$ discussed in Section 2.1 was defined by equation ({\ref{a1def})
to be the Seiberg-Witten period $z_1$
rescaled by a factor $1/\sqrt{\delta}$. Since $z_1$ gives the physical
BPS mass (\ref{MBPS}), we see that the limit involves a rescaling of the
mass scales. The precise formulation of the
conjecture is thus that the correlators of
the gauge theory have a finite limit when expressed in terms of rescaled
space-time variables
\begin{equation}
\label{xnew}
x' = \sqrt{\delta}\, x.
\end{equation}
For more general critical points, the scaling involves different
positive powers of $\delta$, as explained in Section 3, but the idea 
remains the same.

The fact that the double scaling limit is a low energy limit shows
that the study of the Seiberg-Witten periods, which determine the leading
terms of the effective action in a derivative expansion, goes a long way
toward proving that all the other correlators must have a well-defined
limit as well. The reason is that all the non-zero correlators at large
distance are determined by this effective action.  
However, there are two important subtleties that must be taken into 
account. The first one is that in general, the standard local abelian
effective action is not valid near the critical points, because we can 
have light dyons which are not local with respect to each other \cite{AD}.
This argument does not apply to
the simplest critical point discussed in Section 2.1, for which the low
energy theory is a simple abelian gauge theory coupled to a single
charged hypermultiplet. In that case, it may be tempting to incorrectly
identify the double scaled gauge theory with this simple abelian theory.
A way to understand that this cannot possibly be the right answer
is to note that an abelian gauge theory is not a well-defined quantum
theory on all scales, whereas the double scaled theory, which is obtained
by a consistent limiting procedure from a non-abelian gauge theory, must
be a well-defined quantum theory on all scales. What we must get is a
modified version of an abelian gauge theory with a well-defined UV fixed
point. Interestingly, it is actually possible to get different UV
regularizations for a given critical point, as we will see at the end of
Section 3. 

The above discussion may seem a little bit abstract, but all this
physics can be studied very concretely in the context of
two dimensional non-linear $\sigma$ models \cite{fer3}.
Typically, the low energy CFT is then an element of the minimal
series. The double scaled theory can be the associated
Landau-Ginzburg quantum field theory, but also a more general field
theory describing the same physics in the IR but different in the UV
\cite{fer3}.
\subsubsection{The continuum string theory}
The most difficult part of the conjecture is to understand that
a continuum string theory is defined by the 
double scaling limit. Even in the $D\leq 1$ cases studied in
\cite{BK,review}, it is not possible to give a rigorous proof, and our four 
dimensional examples come with several additional subtleties. 

The standard heuristic argument is based on the analysis of the 
$1/N$ expansion of the partition function (\ref{Zexp}), or more 
generally of any amplitude $\cal A$. By taking into account
the contributions from Feynman diagrams only (which obviously are the only 
one that survive in the large $N$ limit in the case of simple zero
dimensional integrals), a suitably normalized amplitude can be expanded,
\begin{equation}
\label{amplexp}
{\cal A} = \sum_{h\geq 0} N^{2-2h}\, A_{h}(g)\, ,
\end{equation}
with
\begin{equation}
\label{amplexp2}
A_{h}(g) = \sum_{n\geq 0}A_{h,n}\, g^{2n}.
\end{equation}
The asymptotic estimate of the coefficients $A_{h,n}$ is 
\begin{equation}
\label{Ahna}
A_{h,n} \mathop{\propto}\limits _{n\rightarrow\infty} a_{h}\, 
n^{\gamma_{h}-3} g_{\rm c}^{-2n}
\end{equation}
for some $a_h$ (for a detailed account of this property and related topics,
see \cite{bes}),
which shows that the series for $A_{h}(g)$ has a finite radius of 
convergence $\gc$ (independent of $h$) and
\begin{equation}
\label{Ahlim}
A_{h}(g) \mathop{\propto}\limits _{g\rightarrow\gc} {a_h\over 
\left(1-(g/\gc)^2\right)^{\gamma_{h}-2}}\cdotp
\end{equation}
When $g\rightarrow g_{\rm c}$, the terms at very large $n$ dominate in the 
sum (\ref{amplexp2}). They are associated with discretized Riemann surfaces of 
area $A=n a^{2}$, where $a$ sets the length scales on the discretized 
world sheet. From (\ref{Ahna}) and (\ref{amplexp2}) 
we see that the contribution of a surface of area $A$ is 
proportional to $\exp (-\lambda A)$, with a renormalized world sheet
cosmological constant $\lambda$ given by
\begin{equation}
\label{wscc}
\lambda = { 2\ln (\gc /g)\over a^{2}} = {\delta_g\over a^{2}}\cdotp
\end{equation}
The continuum limit corresponds to letting the world sheet UV cutoff $1/a$ 
goes to infinity while $\delta_g \rightarrow 0$, keeping fixed $\lambda$.
A scaling limit for which 
contributions from all genera survive can be defined because the exponent 
$\gamma_{h}$ in (\ref{Ahna}) satisfies $2-\gamma_{h} = (2-\gamma_{\rm 
str})(1-h)$ for some $h$-independent string susceptibility $\gamma_{\rm str}$.
By taking
\begin{equation}
\label{dsl}
N\rightarrow\infty\, , \quad \delta_{g}\rightarrow 0\, ,\quad
N\delta_g^{1-\gamma_{\rm str}/2} = {\rm constant,}
\end{equation}
the amplitude (\ref{amplexp}) reduces to an amplitude for an oriented 
closed string theory,
\begin{equation}
\label{Aclosed}
{\cal A} = \sum_{h\geq 0} A_{h}\, \kappa^{2h-2}.
\end{equation}
The scaling (\ref{dsl}) has the interesting consequence that the classical 
string coupling $1/N$ is renormalized to a dimensionfull coupling
\begin{equation}
\label{csc}
g_{\rm str} = {1\over N a^{2-\gamma_{\rm str}}}\cvp
\end{equation}
of world sheet dimension $2-\gamma_{\rm str}$.
The dimensionless coupling $\kappa$ in terms of which the string 
amplitudes are expanded is 
\begin{equation}
\label{kappadef2}
\kappa = g_{\rm str}\lambda^{\gamma_{\rm str}-1/2}.
\end{equation}
In these simple non-critical string theories, a variation of 
the string coupling $g_{\rm str}$ can thus be exactly compensated for by a
variation of a world sheet coupling $\lambda$.

In our four dimensional examples, several of 
the above statements need to be refined, but we are going to argue that 
the general ideas remain valid. The first obvious difference is that the 
't~Hooft coupling $g$ is no longer a free parameter. 
Ordinary perturbation theory is defined by renormalizing the space-time
theory at some scale $\mu$, and the 't Hooft coupling 
is then fixed and equal to
\begin{equation}
\label{RG}
{1\over g^{2}(\mu/\Lambda)} = 
{1\over 4\pi^{2}}\, \ln {\mu\over\Lambda}\cdotp
\end{equation}
Any choice $\mu>\Lambda$ is consistent and yields a real coupling.
For simplicity, let us consider
a case where we vary only one Higgs vev parameter $v$ corresponding to the
most relevant deformation from the critical point. The gauge theory
depends on the dimensionless ratio $r$ defined by (\ref{rdef}).
The expansion (\ref{amplexp2}) is then more precisely written
\begin{equation}
\label{renah}
A_h(r) = \sum_{n\geq 0} A_{h,n}(\mu /v)\, g^{2n}(\mu /\Lambda).
\end{equation}
Note that the $\Lambda$-dependence only appears through the coupling $g$ 
in perturbation theory.
We thus have a different expansion for each different choices of $\mu$. 
However, the critical properties of the series, such as (\ref{Ahna}) or 
(\ref{Ahlim}), cannot depend of the choice of $\mu$, because the physical 
amplitude $\cal A$ is $\mu$-independent. This implies that the 
$A_h$ themselves are $\mu$-independent, because the $\beta$ function 
for the 't~Hooft coupling does not depend on $N$ in our theories. In other 
words, the RG equations insure that the $n\rightarrow\infty$ asymptotics 
of the coefficients $A_{h,n}(\mu /v)$ are independent of $\mu$. We can be 
more precise: if the radius of convergence of (\ref{renah}) for the 
special choice $\mu =v$ 
is $\gc$ ($\gc$ is actually infinite in all the cases we have studied),
then the radius of convergence of (\ref{renah}) for an 
arbitrary $\mu>\Lambda$ is
\begin{equation}
\label{gcform}
\gc^2 (\mu /v) = {\gc^2\over \displaystyle 1+ {\gc^2\over 4\pi^2}\,
|\ln {\mu\over v}\,\,|}\cdotp
\end{equation}
This is demonstrated by noting that the series at general $\mu$ is obtained 
by substituting
\begin{equation}
\label{newrensc}
g^2 = {g^2(\mu /\Lambda)\over 1+ \displaystyle
{g^2(\mu /\Lambda)\over 4\pi^2}\, \ln {v\over\mu}}
\end{equation}
in the series for $\mu=v$. This implies that 
$\gc^2(\mu /v) \geq \gc^2 /\bigl(1+ (\gc^2/4\pi^2) |\ln(\mu /v)|\bigr)$, a
strict inequality being possible mathematically. However, since the 
$\mu$-independent quantities
$A_{h}$ diverge for $r=\rc=v_{\rm c}/\Lambda$ such that 
$g^{2}(\rc) = \gc^{2}$, we must have $g^{2}(\mu /\Lambda) = \gc^{2}(\mu 
/v_{\rm c})$ for all $\mu$, which proves (\ref{gcform}). 
When $r\rightarrow\rc$, $\gc^{2}-g^{2}\propto r-\rc = \delta$, and the 
$\mu$-independent world sheet cosmological constant is given by
$\lambda = \delta /a^{2}$.
The continuum limit is then $\delta\rightarrow 0$ at fixed $\lambda$ and
the double scaling limit is as in (\ref{dsl}) with $\delta_{g}$ replaced 
by $\delta$.

Let us note that the simplifying assumptions used in the above discussion, in 
particular the fact that the $\beta$ function is $N$-independent, are by no 
means necessary conditions for double scaling limits to exist. If the 
$\beta$ function is $N$-dependent, as occurs in less supersymmetric 
situations and in particular in the two dimensional models studied in 
\cite{fer2}, then double scaling limits must involve a mixing between 
different orders in $1/N$, and in particular the naive scaling (\ref{dsl}) 
is corrected by logarithms. Even in our ${\cal N}=2$ examples, this 
complication can occur, because simple scaling laws like (\ref{Ahlim}) 
can be corrected by logarithmic terms, as we will see in Section 3.
\subsubsection{A theory of open strings}
Our preceding discussion focused exclusively on the contributions coming from 
Feynman diagrams. However, one of the main results
of \cite{fer} was to show that, 
in ${\cal N}=2$ super Yang-Mills, there is another type of contribution 
that plays an equally important r\^ole, the fractional instantons. These 
fractional instantons were interpreted as coming from the quantum
disintegration of the large instantons that are present semi-classically.
Remarkably, these fractional instantons generate a series in $1/N$, which 
fits perfectly in a string picture, provided one introduces open strings in 
addition to closed strings. We do not have a graphical representation for 
these open strings, analogous to the double line representation of
Feynman diagrams for the closed strings. However, our conjecture is that 
the double scaling limit corresponds to a continuum limit for both open and 
closed strings. This is particularly significant, because the 
Seiberg-Witten periods that we will study in Section 3
have only a trivial one-loop contribution from the Feynman diagrams.

The presence of open strings might be interpreted by considering that the 
dual field theories are not confining. However, it is not clear whether our 
open strings can live in the bulk of space-time, or are actually stuck 
on some D-brane. The latter possibility seems more likely, because the 
open strings in ${\cal N}=2$ super Yang-Mills \cite{fer} could be
attached to the inflating branes that make the supergravity 
background well-defined through the enhan\c con mechanism \cite{pol}.
\subsubsection{Non-perturbative string theory}
An important feature of our double scaling limits is that they always 
provide a non-perturbative definition of the string theories, thus 
bypassing a fundamental limitation of the $c\leq 1$ matrix models. The way 
this non-perturbative definition is obtained is in some sense 
conservative, since it is based on the use of gauge theory path integrals. 
Note however that our string theories are not dual to gauge theories, but 
to other types of field theories. Our point is that
the gauge theory path integrals {\it in the vicinity of critical points}
automatically generate string theories.

Good simplified models for 
the $c\leq 1$ matrix models are the vector models, for which one 
integrates 
over $N$-vectors instead of $N\times N$-matrices. Double scaling limits 
can be defined in that context, and they generate
continuum theories of polymers \cite{polym,ZJM}. 
For the same reasons as in the matrix models, the double scaling limits 
only give a perturbative definition in the cases that are not Borel summable.
We have argued that for the matrix integrals, this problem is
solved by considering gauge theories integrals. The similar idea for
vector models is to consider non-linear $\sigma$ models integrals.
In bosonic cases, 
one can then explicitly generate non-Borel summable partition functions 
that are nevertheless non-perturbatively defined by the original vector 
integrals \cite{fer4}.

Another related interesting feature of our conjecture is that 
it makes a direct link between the existence of CFT in four dimensions
and the fact that 4D field theories can be dual to string theories.
The modern starting point for the field theory/string theory duality is 
indeed a correspondence between a conformal gauge theory and a string 
theory \cite{malda}, 
but this is an a priori surprising feature because the string 
picture most naturally emerges in confining theories. The fact that a 
string description of CFT is possible comes from the rather non-trivial 
observation that the varying
string tension is set by the Liouville coordinate \cite{wall}. In our 
approach, the string theory is directly generated by the Feynman diagrams 
of a gauge theory at some critical point, without refering to flux 
tubes or confinement. 

One may wonder whether a correspondence between conformal {\it gauge} 
theories and string theories could be derived within our approach. It is 
well-known that such critical points can appear on the moduli space of 
various ${\cal N}=2$ supersymmetric gauge theories. For example, the scale
invariant ${\rm SU}(N)$ theory with 
$N_{\rm f}=2N$ quark hypermultiplets in the fundamenetal representation, 
and various related cousins, are discussed in \cite{arg}. In those cases, 
it is very natural to obtain a theory with both open and closed strings, 
because the large $N$ limit is also a large $N_{\rm f}$ limit. This 
suggests that the discussion of the present paper could be generalized.
As for the ${\cal N}=4$ theory, which is not believed to 
contain open strings, an explicit construction along the lines of 
\cite{arg} does not seem to have appeared. It would be extremely 
interesting to investigate this particular theory in more details.
\section{Exact amplitudes}
Let us now compute explicitly the double scaled Seiberg-Witten period 
integrals $z$ for a variety of critical points. The basic formula for $z$, 
as reviewed in \cite{fer}, is \cite{SW,sun}
\begin{equation}
\label{zdef}
z_{\alpha} = {1\over 2i\pi}\, \oint_{\alpha} {x dp\over y}\, \cvp
\end{equation}
where $\alpha$ is a cycle in the integer homology of the genus $N-1$
hyperelliptic curve
\begin{equation}
\label{curve}
{\cal C}:\ y^{2} = q(x)^{2}= 
p(x)^{2} - 1/r^{2N} = \prod_{i=1}^{N}(x-\phi_{i})^{2} - 1/r^{2N}.
\end{equation}
The $\phi_{i}$s are related to the Higgs vevs moduli and satisfy 
$\sum_{i=1}^{N}\phi_{i}=0$. We choose the contours $\alpha$ to be the 
vanishing cycles at the critical point under study. If those cycles have 
a non-zero intersection form, the low energy theory is an interacting CFT.
As explained in \cite{fer}, 
from the point of view of the large $N$ limit, it is useful to distinguish 
between two classes of singularities depending on whether they occur when a 
classical root of $q$ melts with the enhan\c con (first class)
or when two disconnected pieces of the enhan\c con collide with each other 
(second class). The low energy space-time
CFT does not depend upon the singularity 
being first class or second class, but the double scaled theory, the 
scaling, and the world sheet theory
do depend on the class. In all the examples we have studied, the 
string susceptibility turns out to be
\begin{equation}
\label{gammastr}
\gamma_{\rm str} =0,
\end{equation}
but the simple scaling (\ref{dsl}) can be corrected by logarithms for the 
first class, in a way reminiscent of the $c=1$ matrix model 
\cite{c1}. In the following, we discuss an $A_{n-1}$ series of second 
class singularities generalizing the example $A_{1}$ of Section 2.1,
and then we present the case of the simplest first class singularity.
\subsection{An $A_{n-1}$ series of multicritical points}
We can get a second class $A_{n-1}$ critical point by choosing $N$ to be a 
multiple of $n$ and the polynomial $p$ entering in (\ref{curve}) to be
\begin{equation}
\label{pn}
p(x) = \biggl( x^{n} + \sum_{k=2}^{n-1} u_{n-k} x^{n-k} + 1 \biggr)^{N/n}.
\end{equation}
The scaling limit is universal and does not depend of the precise way the 
second class critical point is embedded in the original gauge theory, so 
the special choice (\ref{pn}) is not restrictive. The parameters $u_{k}$, 
$1\leq k\leq n-2$, and 
\begin{equation}
\label{u0def}
u_{0}= 1-1/r^{n} = \delta\, ,
\end{equation}
correspond to ${\cal N}=2$ preserving relevant deformations around the 
critical point at $u_{k}=0$. Their space-time anomalous dimensions can
be calculated by noting that the curve (\ref{curve}) takes the form
\begin{equation}
\label{singcurve}
y^{2} \approx {2N\over n}\, \Bigl( x^{n} + \sum_{k=2}^{n} u_{n-k}x^{n-k}\Bigr)
\end{equation}
in the vicinity of the singularity and by using the fact that the 
Seiberg-Witten periods are of dimension one. The result is
\begin{equation}
\label{stdim}
[u_{k}] = {2(n-k)\over n+2}\, \cvp\qquad 0\leq k\leq n-2.
\end{equation}
The most relevant operator couples to $u_{0}$. This series of critical 
points corresponds to the $M_{n}^{0}$ SCFTs discussed in \cite{SCFT}.

To understand how the double scaling limit works, let us write the 
polynomial $p$ as
\begin{equation}
\label{plog}
p(x) = \exp \Biggl[ {N\over n}\, \ln \Bigl( 1+\sum_{k=1}^{n-2} u_{k}x^{k} + 
x^{n}\Bigr)\Biggr]
\end{equation}
and change the variable to
\begin{equation}
\label{wdef}
w = N^{1/n}\, x.
\end{equation}
If we take
\begin{equation}
\label{sca1a}
N\rightarrow\infty\, ,\quad u_{k}\rightarrow 0\, ,\quad
t_{k} = N^{1-k/n}\,u_{k} = {\rm constant}\, ,
\end{equation}
then $p$ has a finite limit $\exp \bigl(t(w)/n\bigr)$ with
\begin{equation}
\label{tpoldef}
t(w) = \sum_{k=1}^{n-2}t_{k}w^{k} + w^{n}.
\end{equation}
If, in addition to (\ref{sca1a}), we take
\begin{equation}
\label{sca1b}
N\rightarrow\infty\, ,\quad\delta\rightarrow 0\, ,\quad
t_{0} \mathop{=}\limits ^{\rm def} N\delta = {\rm constant},
\end{equation}
then the rescaled amplitude $N^{1/n}z_{\alpha}$ will also have a finite limit,
\begin{equation}
\label{zlimit}
N^{1/n}\, z_{\alpha} \longrightarrow {1\over 2i\pi n}\,\oint_{\alpha}
{\displaystyle w t'(w)\over\displaystyle\sqrt{1-e^{-2(t(w)+ t_{0})/n}}}\, 
dw.
\end{equation}
The renormalized amplitude, which is finite in the continuum limit, is 
defined to be
\begin{equation}
\label{adefg}
{\cal A}_{\alpha} = {i\pi z_{\alpha}\over\delta^{1/n}}\cdotp
\end{equation}
The scalings (\ref{sca1a}) and (\ref{sca1b}) relate the 
{\it world-sheet} dimensions (not to be confused with the {\it space-time}
dimensions (\ref{stdim})) of the renormalized parameters $t_{k}$,
$\Delta_{k} = (1-k/n)\Delta_{0}$. By identifying the most relevant 
parameter $t_{0}$ with the cosmological constant, we obtain
\begin{equation}
\label{wsdim}
\Delta_{k} = 2-{2k\over n}\cdotp
\end{equation}
This formula suggests that the $t_{k}$s couple to bulk world sheet 
operators ${\cal O}_{k}$ of dimension $2k/n$.
Similarly, the scaling (\ref{adefg}) shows that ${\cal A}_{\alpha}$ is of 
dimension $2/n$.

It is convenient to introduce the new variables and polynomial
\begin{equation}
\label{newvar}
\tau_{k} = (2/n)^{1-k/n}\, t_{k}\, ,\quad u= (2/n)^{1/n} w\, ,\quad
T(u) = \sum_{k=0}^{n-2}\tau_{k}u^{k} + u^{n} ,
\end{equation}
in terms of which
\begin{equation}
\label{finA}
{\cal A}_{\alpha} = {1\over 4\tau_{0}^{1/n}}\, \oint_{\alpha}
{\displaystyle u T'(u)\over\displaystyle\sqrt{1-e^{-T(u)}}}\, du .
\end{equation}
The contour $\alpha$ encircles any two roots of the polynomial $T$. Equation
(\ref{finA}) gives our basic formula for the string theory amplitudes 
corresponding to the $A_{n}$ multicritical points.
Let us look in more details at the special case where only the most 
relevant operator is turned on. It is then natural to introduce the 
coupling $\kappa = 1/\tau_{0}$, to rescale 
$u\rightarrow \kappa^{-1/n}u$, and to consider the contours $\alpha_{jk}$ 
encircling the roots
$u_{j}=\exp (i\pi (1 + 2j)/n)$ and $u_{k}$. A straightforward 
calculation then yields
\begin{equation}
\label{Ai1}
{\cal A}_{\alpha_{jk}} = {n\over 4\kappa}\,\oint_{\alpha_{jk}}
{\displaystyle u^{n}\, 
du\over\displaystyle\sqrt{1-e^{-(1+u^{n})/\kappa}}} =
e^{i\pi (j+k+1)/n}\sin\bigl(\pi(j-k)/n\bigr)\, {\cal I}_{n}(\kappa)
\end{equation}
where
\begin{equation}
\label{Indef}
{\cal I}_{n}(\kappa) = {n\over (n+1)\kappa} + \int_{0}^{1/\kappa}
\Bigl( {\displaystyle1\over\displaystyle\sqrt{1-e^{-x}}}-1 \Bigr) 
(1-\kappa x)^{1/n}dx.
\end{equation}
The asymptotic expansion of ${\cal I}_{n}(\kappa)$ can be obtained by 
noting that when $x\sim 1/\kappa$, $1/\sqrt{1-\exp(-x)} -1$ is 
exponentially small. We thus have
\begin{equation}
\label{Inas}
{\cal I}_{n}(\kappa) = {n\over (n+1)\kappa} + \sum_{k=0}^{K}
{\Gamma(k-1/n)\over\Gamma(-1/n)\Gamma(k+1)} I_{k}\, \kappa^{k} + {\cal 
O}(\kappa^{K+1})
\end{equation}
with
\begin{equation}
\label{Idef2}
I_{k} = \int_{0}^{\infty}\Bigl( {\displaystyle 1\over\displaystyle\sqrt{1
-e^{-x}}} - 1 \Bigr) x^{k}\, dx = {(-1)^{k+1}\over k+1}\,
\int_{0}^{1}{\left( \ln(1-t) \right)^{k+1}\over 2t^{3/2}}\, dt.
\end{equation}
The first integrals $I_{k}$ can be calculated by expanding the logarithm 
in powers of $t$,
\begin{equation}
\label{expik}
I_{0}=2\ln2\, ,\quad I_{1} = {\pi^{2}\over 6} - 2 (\ln 2)^{2}\, ,\quad
I_{2} = {8 (\ln 2)^{2}\over 3} - {2 \pi^{2}\ln 2\over 3} + 4 \zeta(3)\, ,
\ {\rm etc\ldots}
\end{equation}
In particular, we recover the expansion (\ref{a1p}).
\subsection{Other scalings}
We are now going to study the simplest first class singularity. More 
general cases, including mixed examples where first class and second class 
singularities collide, can be studied along the same lines. Let us choose
\begin{equation}
\label{p2nd}
p(x) = (x+1/N)^{N-1} (x-1+1/N).
\end{equation}
The first class singularity occurs at
\begin{equation}
\label{crit1st}
r=\rc = {N\over (N-1)^{1-1/N}}
\end{equation}
when the two positive real roots $x_{1}$ and $x_{-}>x_{1}$
of $p(x)+1/r^{N}$, that exist for $r>\rc$, coincide (see Section 4.1 of
\cite{fer}).
\subsubsection{Elementary analysis}
The large $N$ expansion of the Seiberg-Witten period
\begin{equation}
\label{SW1st}
z = {1\over i\pi}\, \int_{x_{1}}^{x_{-}} {\displaystyle xp'(x)\over
\displaystyle\sqrt{p(x)^{2} - 1/r^{2N}}}
\end{equation}
was evaluated up to terms of order $1/N^{3}$ in \cite{fer}, equation (59),
\begin{eqnarray}
\label{zzz}
&&\hskip -1cm
{i\pi z\over N} = -{r-1\over r} + \ln r + {1\over N}\, \left(
-\ln r + {r-1\over r}\, \ln 2 + {(r-1)\ln(r-1)\over r}\right) \\
\nonumber
&& \hskip 1cm
+{1\over N^2 r}\, \left( {1\over 2}\, \bigl( \ln(r-1) \bigr)^2 + 
(\ln 2)\ln(r-1) + {1\over 2}\, (\ln 2)^2 - {\pi^2\over 24} \right)
+ {\cal O}\bigl( 1/N^3 \bigr).\\ \nonumber
\end{eqnarray}
An important qualitative difference with the expansion at the
second class critical points (as in (\ref{z1}) for example) 
is that there are logarithmic singularities when $r\rightarrow 1$, 
in addition to the simple 
power-like divergences. It is then impossible to find a simple scaling of 
the form (\ref{dsl}). However, let us consider the renormalized amplitude
\begin{equation}
\label{adef1stclass}
{\cal A} = i\pi N z
\end{equation}
and the modified scaling
\begin{equation}
\label{sca1stclass}
N\rightarrow\infty \, ,\quad \delta = r-1\rightarrow 0\, ,\quad
N\delta - \ln N = {\rm constant} = 1/\kappa' .
\end{equation}
Then the leading term in (\ref{zzz}) gives $1/(2\kappa'^{2})$ plus divergent
$(\ln N)/\kappa'$ and $(\ln N)^{2}$ terms. The first divergence is 
actually exactly canceled by taking into account the $1/N$ term in 
(\ref{zzz}). All the divergences at order $\kappa'^{0}$, which come from 
all three terms in (\ref{zzz}), turn out to cancel likewise, yielding 
the finite expansion
\begin{equation}
\label{1ex}
{\cal A} = {1\over 2\kappa'^{2}} - {\ln(e\kappa' /2)\over\kappa'} + 
{(\ln\kappa')^{2}\over 2} - (\ln 2)\ln\kappa' + {(\ln 2)^{2}\over 2} - 
{\pi^{2}\over 24} + \cdots
\end{equation}
We will show below that these cancellations actually work to all orders 
and beyond. The scaling (\ref{adef1stclass}) 
then shows that $\cal A$ has world sheet dimension two.  
The expansion (\ref{1ex}) can be put in a more familiar form by 
introducing a new coupling $\kappa$ defined by
\begin{equation}
\label{kapdef1}
1/\kappa' =1/\kappa + \ln\kappa
\end{equation}
and in terms of which
\begin{equation}
\label{goodexp}
{\cal A} = {1\over 2\kappa^{2}} + {\ln 2 -1\over\kappa} + {1\over 2}(\ln 
2)^{2} - {\pi^{2}\over 24} + {\cal O}(\kappa).
\end{equation}
\subsubsection{The full amplitude}
By changing the variable to $z=N(1/(x+1/N) -1)$ in (\ref{SW1st}) we get
\begin{equation}
\label{aint}
{\cal A} = N \int_{z_{-}}^{z_{1}}\! dz\,  {1-z\over z(1+z/N)^{2}}
{\bigl[1-1/N -z/N^{2}\bigr]\bigl[1+z/(N(1-z))\bigr]
\over\displaystyle\sqrt{1-N^{2}(1+z/N)^{2N}/(z^{2}r^{2N})}}\, \cdotp
\end{equation}
A complication with respect to the cases studied in Section 3.1 stems from 
the fact that $\cal A$ comes with a factor of $N$ in front of the integral.
To show that the scaling (\ref{sca1stclass}) yields a finite limit, we thus 
have to check that the leading term is actually zero. The finite string 
amplitude is then extracted by taking into account subleading terms in the 
integrand and in $z_{-}$ and $z_{1}$. Up to terms that trivially 
go to zero, we obtain
\begin{equation}
\label{aint2}
{\cal A} = {\cal A}_{a} + {\cal A}_{b}
\end{equation}
with
\begin{eqnarray}
\label{aint3}
&&{\cal A}_{a} = \int_{Z_{-}}^{Z_{1}}\! dz\, {2z^{2}-1\over z}\,
{1\over\sqrt{1-e^{2(z-1/\kappa)}/(z\kappa)^{2}}}\, \cvp\label{Aa}\\
&&{\cal A}_{b} = N\int_{\hat Z_{-}}^{\hat Z_{1}}\! dz\, {1-z\over z}\,
{1\over\sqrt{1-(1-z^{2}/N)e^{2(z-1/\kappa)}/(z\kappa)^{2}}}\,
\cdotp \label{Ab}\\ \nonumber
\end{eqnarray}
The boundaries of the integration region for ${\cal A}_{a}$,
$Z_{-}$ and $Z_{1}>Z_-$, are the two positive solutions of
\begin{equation}
\label{fdef}
f(z) = ze^{-z} = f(1/\kappa) \mathop{=}\limits ^{\rm def}\rho .
\end{equation}
It is useful to introduce the inverses of the function $f$,
\begin{equation}
\label{invf}
x=f(z) \Longleftrightarrow\left\{\matrix{ z = & \hskip -.9cm W_{-}(x)\ 
{\rm for}\ z\in [0,1]\cr 
z= & \hskip -.3cm W_{1}(x)\ {\rm for}\ z\in [1,+\infty[\, .\cr}\right.
\end{equation}
One then has
\begin{equation}
\label{bornes}
Z_{-}=W_{-}(\rho)\, ,\quad Z_{1} = W_{1}(\rho) = 1/\kappa\, .
\end{equation}
On the other hand, $\hat Z_{-}$ and $\hat Z_{1}>\hat Z_{-}$ 
are solutions of
\begin{equation}
\label{fd2}
f(z) = \Bigl( 1-{z^{2}\over 2N}\Bigr) \rho\, ,
\end{equation}
and in particular
\begin{equation}
\label{zch}
f(\hat Z_{-}) = \Bigl( 1-{Z_{-}^{2}\over 2N} \Bigr)\rho +{\cal O}(1/N^{2})\, ,
\quad f(\hat Z_{1})=\Bigl( 1-{Z_{1}^{2}\over 2N} \Bigr)\rho + 
{\cal O}(1/N^{2})\, .
\end{equation}
Let us now change the variable in the integral (\ref{Ab}) from $z$ 
to $x=f(z)$. We get
\begin{eqnarray}
\label{A7}
&& {\cal A}_{b}^{<}
= N\int_{f(\hat Z_{-})}^{1/e} {dx\over x}
{1\over\displaystyle
\sqrt{1- (1-W_{-}^{2}(x)/N)\rho^{2}/x^{2}}}\, \cvp\label{A7a}\\
&& {\cal A}_{b}^{>} 
= N\int_{1/e}^{f(\hat Z_{1})} {dx\over x}
{1\over\displaystyle
\sqrt{1- (1-W_{1}^{2}(x)/N)\rho^{2}/x^{2}}}\, \cdotp\label{A7b}\\ \nonumber
\end{eqnarray}
The large $N$ limits of these two integrals are similar. For example, to 
study (\ref{A7a}), we substitute
$x\rightarrow f(\hat Z_{-}) + \rho (x-1)$, and expand the integrand 
at large $N$ by carefully taking into account (\ref{zch}). The result is
\begin{equation}
\label{A8}
{\cal A}_{b}^{<} = 
-{e\rho Z_{-}^{2}\over 2\sqrt{1-e^{2}\rho^{2}}}+
N\int_{1}^{{1\over e\rho}} {dx\over\sqrt{x^{2}-1}}
+{1\over 2}\int_{1}^{{1\over e\rho}} 
{dx\over (x^{2}-1)^{3/2}}
\left( x Z_{-}^{2} - W_{-}^{2}(\rho x)\right) \cdotp
\end{equation}
The formula for ${\cal A}_{b}^{>}$ is the same except for a global 
sign change and the replacement of $W_{-}$ and $Z_{-}$ by $W_{1}$ and 
$Z_{1}$ respectively. The term proportional to $N$ in thus finally
\begin{equation}
\label{zeroN}
N \left( \int_{1}^{{1\over e\rho}}\!\!\! + \int_{{1\over e\rho}}^{1} \right)
{dx\over\sqrt{x^{2}-1}} = 0
\end{equation}
which shows that the amplitude indeed has a finite limit in the scaling 
(\ref{sca1stclass}),
\begin{equation}
\label{fin}
{\cal A}_{b} = 
{e\rho\, (Z_{1}^{2} - Z_{-}^{2})\over 2\sqrt{1-e^{2}\rho^{2}}}
+{1\over 2}\int_{1}^{{1\over e\rho}}\! {dx\over (x^{2}-1)^{3/2}}
\left( W_{1}^{2}(\rho x) - W_{-}^{2}(\rho x) - x 
(Z_{1}^{2}-Z_{-}^{2})\right)\, \cdotp
\end{equation}
In the original variables, we finally end up with
\begin{eqnarray}
\label{fin2}
&& \hskip -1.1cm {\cal A} =  
{e\rho\, (Z_{1}^{2} - Z_{-}^{2})\over 2\sqrt{1-e^{2}\rho^{2}}} +
\int_{Z_{-}}^{Z_{1}}\! dz\, {2z^{2}-1\over z}\,
{1\over\sqrt{1-e^{2(z-1/\kappa)}/(z\kappa)^{2}}}\nonumber\\
&&\hskip -.3cm
+ {\rho^{2}\over 2}\int_{Z_{-}}^{1}\! dz\, {z-1\over z^{2}}\,
{\displaystyle (z-Z_{-}^{2} e^{-z}/\rho )\, e^{2z}\over\displaystyle
 (1-\rho^{2} e^{2z}/z^{2})^{3/2}}
+{\rho^{2}\over 2}\int_{1}^{Z_{1}}\! dz\, {z-1\over z^{2}}\,
{\displaystyle (z-Z_{1}^{2} e^{-z}/\rho )\, e^{2z}\over \displaystyle
(1-\rho^{2} e^{2z}/z^{2})^{3/2}}
\,\cdotp\\ \nonumber
\end{eqnarray}
\vfill\eject
\section{Conclusion}

We have generalized the matrix model approach to non-critical string 
theories in less than one dimension \cite{DK,BK} to non-critical string 
theories in four dimensions. The main differences between the two cases is 
that the latter is always non-perturbatively defined and contains open 
strings in addition to closed strings. There are many tantalizing 
questions that we have not addressed. In particular, a standard continuum 
construction of the world sheet theory is lacking. It would be extremely 
interesting to be able to recover some of our results in this more 
familiar context.

\section*{Acknowledgements}

I would like to thank V.I.~Kazakov, I.R.~Klebanov, V.~Periwal, A.M.~Polyakov, 
L.~Rastelli and K.~Skenderis for
stimulating discussions at various stages of this work. This work was 
supported by a Robert H.~Dicke fellowship.

\begin{thebibliography}{99}
%
\bibitem{BK}{\'E.~Br\'ezin and V.A.~Kazakov, \plb{236}{1990}{144},\\
M.R.~Douglas and S.~Shenker, \npb{355}{1990}{635},\\
D.J.~Gross and A.A.~Migdal, \prl{64}{1990}{127}.}
%
\bibitem{tHooft}{G.~'t~Hooft, \npb{72}{1974}{461}.}
%
\bibitem{DK}{F.~David, \npb{257}{1985}{45},\\
V.A.~Kazakov, \plb{150}{1985}{282},\\
J.~Ambj\o rn, B.~Durhuus and J.~Fr\"ohlich, \npb{257}{1985}{433}.}
%
\bibitem{IZ}{\'E.~Br\'ezin, C.~Itzykson, G.~Parisi and J.B.~Zuber,
\cmp{59}{1978}{35}.}
%
\bibitem{review}{P.~Di Francesco, P.~Ginsparg and J.~Zinn-Justin,
\pr{254}{1995}{1},\\
I.R.~Klebanov, {\it String theory in two dimensions,} Trieste spring
school in String Theory and Quantum Gravity 1991, hep-th/9108019,\\
\'E.~Br\'ezin and S.R.~Wadia editors, {\it The large $N$ expansion
in quantum field theory and statistical physics,} World Scientific 1993.}
%
\bibitem{fer}{F.~Ferrari, \npb{612}{2001}{151}.}
%
\bibitem{fer2}{F.~Ferrari, \plb{496}{2000}{212},\\
F.~Ferrari, \jhep{6}{2001}{57}.}
%
\bibitem{Dorey}{N.~Dorey, \jhep{11}{1998}{005}.}
%
\bibitem{fer3}{F.~Ferrari, {\it The large $N$ expansion, 
fractional instantons, and double scaling limits in two 
dimensions,} PUPT-1997, NEIP-01-008, LPTENS-01/11, in preparation.}
%
\bibitem{bes}{D.~Bessis, C.~Itzykson and J.-B.~Zuber, \aam{1}{1980}{109}.}
%
\bibitem{ZJM}{J.~Ambj\o rn, B.~Durhuus, and T.~J\' onsson, 
\plb{244}{1990}{403},\\
J.~Zinn-Justin, \plb{257}{1991}{335},\\
P.~Di Vecchia, M.~Kato, and N.~Ohta, \ijmpa{7}{1992}{1391},\\
G.~Eyal, M.~Moshe, S.~Nishigaki and J.~Zinn-Justin,
\npb{470}{1996}{369},\\
M.~Moshe, {\it Quantum field theory in singular limits,} Les Houches
lecture 1997, hep-th/9812029.}
%
\bibitem{SW}{N.~Seiberg and E.~Witten, \npb{426}{1994}{19}, erratum
{\bf B 430} (1994) 485,\\
N.~Seiberg and E.~Witten, \npb{431}{1994}{484}.}
%
\bibitem{AD}{P.C.~Argyres and M.R.~Douglas, \npb{448}{1995}{93}.}
%
\bibitem{sun}{P.~C.~Argyres and A.~E.~Faraggi,
\prl{\bf 74}{1995}{3931},\\
A.~Klemm, W.~Lerche, S.~Yankielowicz and S.~Theisen,
\plb{344}{1995}{169}.}
%
\bibitem{pol}{C.V.~Johnson, A.W.~Peet and J.~Polchinski, 
\prd{61}{2000}{86001}.}
%
\bibitem{polym}{S. Nishigaki and T. Yoneya, \npb{348}{1991}{787},\\
P. Di Vecchia, M. Kato, and N. Ohta, \npb{357}{1991}{495}.}
%
\bibitem{fer4}{F.~Ferrari, {\it Double scaling limits in non-linear 
$\sigma$ models}, PUPT-1998, NEIP-01-009, in preparation.}
%
\bibitem{malda}{J.~Maldacena, \atmp{2}{1998}{231},\\
S.~Gubser, I.R.~Klebanov and A.M.~Polyakov, \plb{428}{1998}{105},\\
E.~Witten, \atmp{2}{1998}{253}.}
%
\bibitem{wall}{A.M.~Polyakov, {\it Nucl. Phys. Proc. Suppl.} {\bf 68} 
(1998) 1,\\
A.M.~Polyakov, \ijmpa{14}{1999}{645}.}
%
\bibitem{arg}{P.C.~Argyres, \atmp{2}{1998}{293}.}
%
\bibitem{c1}{\'E.~Br\'ezin, V.I.~Kazakov and Al.B.~Zamolodchikov, 
\npb{338}{1990}{673},\\
G.~Parisi, \plb{238}{1990}{209},\\
D.~Gross and M.~Milkovic, \plb{238}{1990}{217},\\
P.~Ginsparg and J.~Zinn-Justin, \plb{240}{1990}{333}.}
%
\bibitem{SCFT}{T.~Eguchi, K.~Hori, K.~Ito and S.-K.~Yang, 
\npb{471}{1996}{430}.}
%
\end{thebibliography}
\end{document}